\documentclass[english]{article}
\usepackage[T1]{fontenc}
\usepackage{geometry}
\usepackage{graphicx}

\makeatother

\providecommand{\tabularnewline}{\\}

\newcommand{\lyxaddress}[1]{
\par {\center #1
\vspace{1.4em}
\noindent\par}
}

\usepackage{babel}
\makeatother
\begin{document}

\title{On two approaches to the building of local models for electron density
based on Irkutsk digizond data.}

\author{Berngardt O.I.}

\maketitle

\lyxaddress{Institute of Solar-Terrestrial Phyisics, SB RAS, Irkutsk,Russia,}

\lyxaddress{E-mail: berng@iszf.irk.ru}

\begin{abstract}
In the paper the step-by-step principles for making local model of
electron density are described. They are based on modulation principle
- electron density dependence on time is a product of a number of
temporal variations caused by solar radiation, magnetic activity,
Earth orientation and unknown additional periodical processes (not
a sum, as they suppose sometimes when making such models). A multiranges
modulation principle is also suggested, that allows automatically
extend the set of parameters by using new ones, obtained by filtration
(or averaging) of basic set of parameters over the time. In the paper
we describe two approaches to the model creation - descriptional and
predictional ones.

To test the approach three different models were created for daily
electron density logarithm using the described principles. We have
used the data of Irkutsk digisonde over the period 2003-2007 years
for testing. It becomes clear that a non-optimal choice of the number
of model parameters could increase prediction error, inspite the error
over the set, used for analysis, will decrease. It is shown that one
year prediction has accuracy about 9-23\% depending on the height,
and the highest error corresponds to the height about 200km. From
the modelling we could also see that with increasing of the height
the number of parameters increases, and this could be caused by inaccuracy
of the model or by not taking additional physical mechanisms into
consideration. 
\end{abstract}

\section{Introduction}

One of the standard approaches to the building of local models is
regression model for electron density \cite{Holt_et_al_2002}:

\begin{equation}
N(r,t)=a(r,t)+b(r,t)f_{10.7}(t)+c(r,t)Ap(t)+d(r,t)f_{10.7}(t)Ap(t)\label{eq:1}\end{equation}

Inspite of the fact that the model is very simple - it is just an
interpolation of arbitrary function (electron density) as a Taylor
sequence using $f_{10.7}(t),Ap(t)$ as small parameters, it gives
a good agreement with the experiment. But it does not depend on electron
density creation mechanisms, and due to this it might have sometimes
a weak prediction accuracy, inspite it could describe the data used
for approximation pretty accurately. In some approaches to increase
the model accuracy one could take into account a dependence on averaged
values of the parameters $f_{10.7}(t),Ap(t)$ too, for example on
$f_{10.7}(t)$, averaged over 81 and 3 days.

In the work I tried to suggest a way to build a regression local model
with some physycal principles included.

\section{Modulation principle as a basis for the model.}

At first let us recall that electron density layers formation is mostly
caused by solar radiation, and the effective radiation wavelength
depend on the height\cite{Hargreaves_1992}. Also, these mechanisms
depend on Earth orientation and cause a number of additional modulation
effects. In the article we will analyze electron density at a selected
height, the other heights could be calculated independently using
the same approach. 

At first let us analyze the modulation effects causing electron density
variations at a selected height.

The first effect is a modulation of solar radiation due to solar activity
changes. These changes could be quantitatively characterized by solar
sunspot number ($W$ - Wolf numbers ) or by $f_{10.7}$ index - solar
radiation at 10.7 cm wavelength. Inspite of ionizing radiation and
$f_{10.7}$ has different wavelengths (1nm and 100nm for D-layer,
1-10nm for E-layer, 10-100nm for F-layer), it is clear that $f_{10.7}$
is a more correct parameter than $W$ due to the fact that it could
be defined more correct.

The second modulation effect is electron density dependence on zenith
angle of the sun as a function of daytime, day of year, longitude
and latitude. This dependence controls the ionization processes and
forms (together with some other mechanisms) the electron density dependence
on height and time, also well-known as Chapmen layer\cite{Chapmen_1931}.
That is why the zenith angle is also well defined parameter that must
be included into the model. In this work we do not analyze hourly
variations of the electron density and due to this we use only position
of the Earth at orbit as a basic parameter, which defines day-to-day
variations of the zenith angle (at given latitude,longitude and height):\begin{equation}
Decl(t)=-0.40915*cos(2\pi(DayNo(t)+8)/365.25)\label{eq:2}\end{equation}

where $DayNo(t)$ - day number within a year.

We do not analyze electron density variations with periods less that
1 day so the usage of this parameter looks correct to us. For building
the model that includes variations with periods less than a day this
parameter should be changed or new parameter should be added to characterize
hourly variations of electron density. 

The third parameter, traditionally included into models, is a magnetic
activity index. Exact dependence of electron density on magnetic activity
is not clear, but we suppose that it also modulates electron density.

The forth modulation parameter includes all the hidden periodics that
was not included into our model by previous 3 parameters. They might
exist and be caused by different periodical changes in the ionization
processes. There is a number of such potential processes, for example
changes of neutral athmosphere, moon effects, etc. We do not know
exact frequencies $\omega_{i}$ and phases $\varphi_{i}$ of these
periodics, so their number, phases and periodics should be found from
data analysis.

Finally the model of electron density could be written in a form:

\begin{equation}
Ne(t)=C_{0}(f_{10.7}(t))*C_{1}(Ap(t))*C_{2}(Decl(t))*\begin{array}{c}
n\\
\Pi\\
i=1\end{array}C_{3+i}(cos(\omega_{i}t+\varphi_{i}))\label{eq:3}\end{equation}

where $C_{n}(x)$ - some positive functions, and $\Pi$ - product
sign.

\section{On choice of the basis functions. }

When positive function is a product of positive modulation functions
(electron density is a positive function by definition) it is useful
to analyze logarithm of this function as a sum of some functions,
each of them defines its own modulation effect independently, i.e.
these modulation functions work as a basis:

\begin{equation}
\begin{array}{l}
log(Ne)=D_{0}(f_{10.7}(t))+D_{1}(Ap(t))+D_{2}(Decl(t))+\begin{array}{c}
n\\
\Sigma\\
i=1\end{array}D_{3+i}(cos(\omega_{i}t+\varphi_{i}))\end{array}\label{eq:4}\end{equation}

where

\begin{equation}
D_{i}(x)=log(C_{i}(x))\label{eq:5}\end{equation}

One of well known applications that uses this approach is the so-called
cepstral analysis \cite{Gonorovskij_1986} - analysis of the logarithm
of the function as a sum of periodical functions. Ideologically our
technique is close to the cepstral analysis, but we use some non-peridical
(but lineary independent) functions in addition to the periodical
functions.

Let us define a structure of the unknown functions $D_{i}$ from some
basic physical principles. 

The structure of the $D_{0}(x)$ is clear: due to the ionising component
of the radiation wavelength differs from characterstic index $f_{10.7}$
wavelength and most of the spectras in a wide region could be approximated
by power low, the electron density could be in first approximation
proportional to the power of the $f_{10.7}$:

\begin{equation}
Ne(t)=C_{01}(f_{10.7}(t))^{C_{02}}\label{eq:6}\end{equation}

where $C_{01},C_{02}$ - some parameters. Inspite of solar spectrum
shape could depend on solar activity level, at first approximation
we could suppose that its shape remains constant and only its amplitude
is changed with changing $f_{10.7}$. This automatically means that
in the absence of the solar radiation the electron density becomes
zero, this suggestion does not contaminate with basic physical principles. 

So the $D_{0}(x)$ becomes:

\begin{equation}
D_{0}(f_{10.7}(t))=C_{01}+C_{02}log(f_{10.7}(t))\label{eq:7}\end{equation}

Lets also suppose that dependence on magnetic activity is also defined
by power low (we do this analogous to the solar activity dependence
and this could not be explained from physical point of view):

\begin{equation}
D_{1}(Ap(t))=C_{11}+C_{12}log(Ap(t))\label{eq:8}\end{equation}

Let us suppose that the dependence on daily variations has an exponential
character:

\begin{equation}
Ne(t)=C_{21}exp(C_{22}Decl(t))\label{eq:9}\end{equation}

due to zenith angle exponentially defines electron density for Chapmen
layer at a selected height \cite{Chapmen_1931}:

\begin{equation}
Ne(t)=Aexp(-B\frac{1}{cos(\chi)})\label{eq:9.1}\end{equation}

or, according to \cite{Shimazaki_1959} by extracting the dependence
on zenith angle at local noon $\chi_{0}$: 

\begin{equation}
Ne(t)=Aexp(-B\frac{cos(\chi_{0})}{cos(\chi)})\label{eq:9.1b}\end{equation}

where \begin{equation}
cos(\chi)=sin(LAT)\cdot sin(Decl(t))+cos(LAT)\cdot cos(Decl(t))\cdot sin((XLT(t)-6.)\cdot0.2617994)\label{eq:9.2}\end{equation}

$XLT(t)$ - local solar time.

So $D_{2}(x)$ could be written (in a first approximation, taking
into account only linearized part of argument in (\ref{eq:9.1b})
) as:

\begin{equation}
D_{2}(Decl(t))=C_{21}+C_{22}Decl(t)\label{eq:10}\end{equation}

Other periodical functions are just linear ones:

\begin{equation}
D_{3+i}(x)=C_{3,1}x\label{eq:11}\end{equation}

By grouping all these together we obtain the following model for electron
density:

\begin{equation}
\begin{array}{r}
Ne_{log}(t)=log(Ne(t))=C+C_{02}log(f_{10.7}(t))+C_{12}log(Ap(t))+\\
+C_{22}Decl(t)+\begin{array}{c}
n\\
\Sigma\\
i=1\end{array}C_{3+i,1}(cos(\omega_{i}t+\varphi_{i}))\end{array}\label{eq:12}\end{equation}

where \begin{equation}
C=C_{01}+C_{11}+C_{21}\label{eq:13}\end{equation}

Actually, the ionosphere is pretty innertial system so its responce
to the disturbancies of different periods could be different. Within
the suggested approach we could suppose that different periods of
disturbancies cause an independent modulation (we will call this multiranges
modulation principle). Let us suppose that in the electron density
time dependence we could find some well-defined response time: 1 day
as Earth rotation period, 3 days as average disturbance period and
30 days as approximate solar rotation/moon rotation period. The choice
of the number and exact values of these periods is not principal for
now and is used for illustration of the method.

Lets define logarithm of electron density $Ne_{log}(t)$ as a sum
of functions with given periods of changes: 

\begin{equation}
Ne_{log}(t)=Ne_{log,<1d}(t)+Ne_{log,[1d,3d]}(t)+Ne_{log,[3d,30d]}(t)+Ne_{log,>30d}(t)\label{eq:14}\end{equation}

The principle of making these functions $Ne_{log,<1d}(t)$, $Ne_{log,[1d,3d]}$,
$Ne_{log,[3d,30d]}(t)$, $Ne_{log,>30d}(t)$ from the source function
$Ne_{log}(t)$ is not important. It is more important that the sum
is exactly $Ne_{log}(t)$ and these functions have necessary time
changes (\ref{eq:14}). One of the most simple methods is using the
time averaging or using filters with rectangular bands for almost
all the functions (1-3days period, 3-30days period and more than 30days
period). The last function (for example with 0-1 days periods) is
defined to the condition (\ref{eq:14}) becomes true.

Each of the components is approximated by the sum (\ref{eq:12}):

\begin{equation}
\begin{array}{r}
Ne_{log,T}(t)=C_{T}+C_{02,T}log(f_{10.7}(t))+C_{12,T}log(Ap(t))+\\
+C_{22,T}Decl(t)+\begin{array}{c}
n\\
\Sigma\\
i=1\end{array}C_{3+i,1,T}(cos(\omega_{i}t+\varphi_{i}))\end{array}\label{eq:15}\end{equation}

where index $T$ defines period ranges of corresponding variations:
<1days, {[}1day,3days], {[}3days,30days], >30days

The resulting model for the electron density logarithm is the following:

\begin{equation}
\begin{array}{r}
Ne_{log}(t)=C+\begin{array}{c}
\\\Sigma\\
T\end{array}C_{02,T}log_{T}(f_{10.7}(t))+\begin{array}{c}
\\\Sigma\\
T\end{array}C_{12,T}log_{T}(Ap(t))+\\
+C_{22}Decl(t)+\begin{array}{c}
N\\
\Sigma\\
i=1\end{array}C_{3+i,1}(cos(\omega_{i}t+\varphi_{i}))\end{array}\label{eq:16}\end{equation}

where the sum is calculated over all the period ranges. The index
$T$ means applying the same filtration technique as used by us for
selecting electron density periods to the argument functions:

\begin{equation}
log(x)=log_{<1d}(x)+log_{[1d,3d]}(x)+log_{[3d,30d]}(x)+log_{>30d}(x)\label{eq:17}\end{equation}

\begin{equation}
Decl(x)=Decl_{<1d}(x)+Decl_{[1d,3d]}(x)+Decl_{[3d,30d]}(x)+Decl_{>30d}(x)\label{eq:18}\end{equation}

So, according to this multiranges modulation principle we should choose
important ionospheric response times and make the extended set of
arguments by filtration of all the indexes with some filtration technique. 

In the paper we will analyze only variations with periods no less
than a day. In this case the model (\ref{eq:16}) becomes the following:

\begin{equation}
\begin{array}{r}
Ne_{log,\geq1d}(t)=C+\begin{array}{c}
\\\Sigma\\
T\end{array}C_{02,T}log_{T}(f_{10.7}(t))+\begin{array}{c}
\\\Sigma\\
T\end{array}C_{12,T}log_{T}(Ap(t))+\\
+C_{22}Decl(t)+\begin{array}{c}
N\\
\Sigma\\
i=1\end{array}C_{3+i,1}cos(\omega_{i}t+\varphi_{i})\end{array}\label{eq:19}\end{equation}

From right and left parts we have removed all the effects with periods
less than a day. 

Shortly, let us describe the model (\ref{eq:19}):

1) According to the filtration rule (\ref{eq:14}) we should average
logarithm of electron density at given height for whole the day. This
corresponds to the geometrical averaging of the electron density with
following calculation of the logarithm. 

2) According to the filtration rule (\ref{eq:14}) we should create
extended set of indexes with period ranges 1 day,1-3days,3-30days
and more than 30days periods from basic set of paramters $f_{10.7}(t),Ap(t)$: 

\begin{equation}
\begin{array}{l}
log_{1d}(f_{10.7}(t)),log_{[1d,3d]}(f_{10.7}(t)),log_{[3d,30d]}(f_{10.7}(t)),log_{>30d}(f_{10.7}(t)),\\
log_{1d}(Ap(t)),log_{[1d,3d]}(Ap(t)),log_{[3d,30d]}(Ap(t)),log_{>30d}(Ap(t)),\\
Decl_{T}(t)\end{array}\label{eq:20}\end{equation}

3) Next step is the calculation of the coefficients $C,C_{02,T},C_{12,T},C_{22}$from
(\ref{eq:19}) and minimizing the functional:

\begin{equation}
\begin{array}{r}
\Omega(\{ C\})=\begin{array}{c}
\\\Sigma\\
t\end{array}(Ne_{log,\geq1d}(t)-C+\begin{array}{c}
\\\Sigma\\
T\end{array}C_{02,T}log_{T}(f_{10.7}(t))+\begin{array}{c}
\\\Sigma\\
T\end{array}C_{12,T}log_{T}(Ap(t))+\\
+C_{22}Decl_{T}(t)+\begin{array}{c}
N\\
\Sigma\\
i=1\end{array}C_{3+i,1}cos(\omega_{i}t+\varphi_{i}))^{2}\end{array}\label{eq:21}\end{equation}

At first step we suppose $N=0$.

4) According to the technique \cite{Kuklin_et_al_2000} we will increase
$N$ by 1 until some stop condition is reached, and repeat actions
from step 3). We determine parameters $C_{3+N,1}$, $\omega_{N}$,
$\varphi_{N}$ for one periodic at each step based on functional (\ref{eq:21})
minimum, using all the previous periodics coefficients $C_{3+i<N,1}$
, $\omega_{i<N}$, $\varphi_{i<N}$and parameters $C,C_{02,T},C_{12,T},C_{22}$
determined at previous steps. 

The only problem for using this approach is to choose correct stop
condition (we could increase $N$ up to infinity). Lets analyze the
problem of stop condition more detailed.

\section{Determination of periodical components - two approaches to the modelling.}

There are at least two close but not similar approaches for solving
this task. Of couse we are ordering parameters by their effectivity
(energy of corresponding model component).

Approach 1, or description aproach: to obtain a maximal accuracy for
description of the data.

It is used most frequently. It is clear, that when using large enough
set of periodical functions, we could approximate almost any function
as accurate as required. The most effective way to demonstrate it
is Fourier transform, that completes this task exactly. That is why
in this approach the number of periodics could not be limited by the
stated condition - we always could add next periodic and obtain higher
accuracy of the model. So usually one could limit number of periodics
by hands (for example - calculate only 3, 5 or 10 periodics) or stop
when amplitude of periodic becomes lower than somehow defined noise
of the data. 

Approach 2, or prediction approach: to obtain a maximal accuracy for
prediction of the data.

In opposite to the description approach, for prediction approach we
should divide dataset into two non-intersecting intervals - analysis
interval and prediction interval. All the model coefficients are defined
from the analysis interval, but stop condition is defined from prediction
interval, to the accuracy at prediction interval would be as effective
as possible. This approach automatically limits number of periodics
by not taking noise components into consideration. It is clear, that
due to data noise, non-stationarity of the data and effects not included
into the model an error will start to increase after the number of
periodics exceeds some value. While the number of periodics less than
this border number, the model remains adequate. When it exceeds this
border number, the model becomes inadequate. There are at least two
ways to calculate this stop condition within prediction approach.
The first way is to stop when first local minimum of the error is
reached at prediction set (absolute stop condition). The second way
is to stop when a local minimum of relation between error on prediction
set and error on analysis set is reached (relative stop condition).
We will use the second variant for analysis of Irkutsk data in prediction
approach.

\section{Local electron density models based on Irkutsk digisonde data.}

For making the model the data from Irkutsk digisonde DPS-4 \cite{Reinisch_1997}
has been analyzed. The data covers more than 4 years, from 2003 till
2007. From the data analysis there was excluded the most magneticaly
disturbed periods with $Ap>100$. The experimental data for electron
density logarithm, averaged by days, are shown at fig.1. 

\includegraphics[scale=0.5,angle=270]{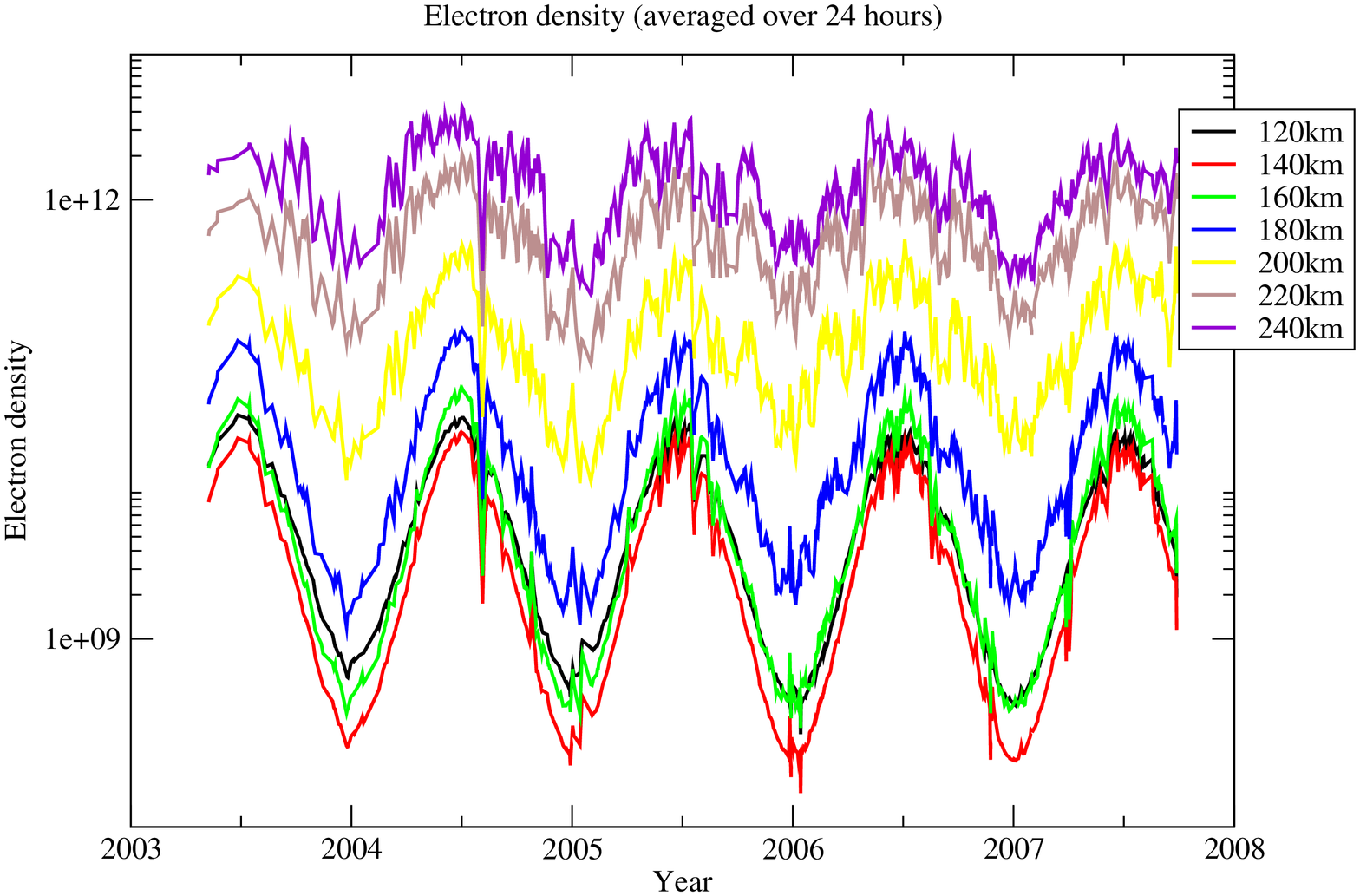}

{\small Fig.1 Logarithm of electron density, averaged over 24 hours
for heights 120-240 km. Irkutsk digisonde data.}{\small \par}

~

Even from the shown data it is clear that the modulation model for
electron density is a very good approximation: as one can see, the
variations with different periods look independent and have amplitudes
independent on time (in average).

For testing the technique we have divided the whole set fo the data
into two subsets: analyze set - 2003-2006 years and prediction set
- 2007 year. 

At fig.2 the comparison between experiment and description model is
shown. The model shown at fig.2 has been built without any additional
periodics ($N=0$). At table 1 the accuracy of this model over the
two different sets - analyze and prediction ones (columns 2-3) is
shown. As one can see, the accuracy for prediction set is a little
bit higher than for analyze set. It looks strange, but could be explained:
absolute values of the electron density decreases with time, as can
be seen from the fig.2 (2007 year is close to solar activity minimum).
The order of density changes is about 30\% and this corresponds to
the observed difference between prediction and analyze sets accuracy.

\includegraphics[scale=0.5,angle=270]{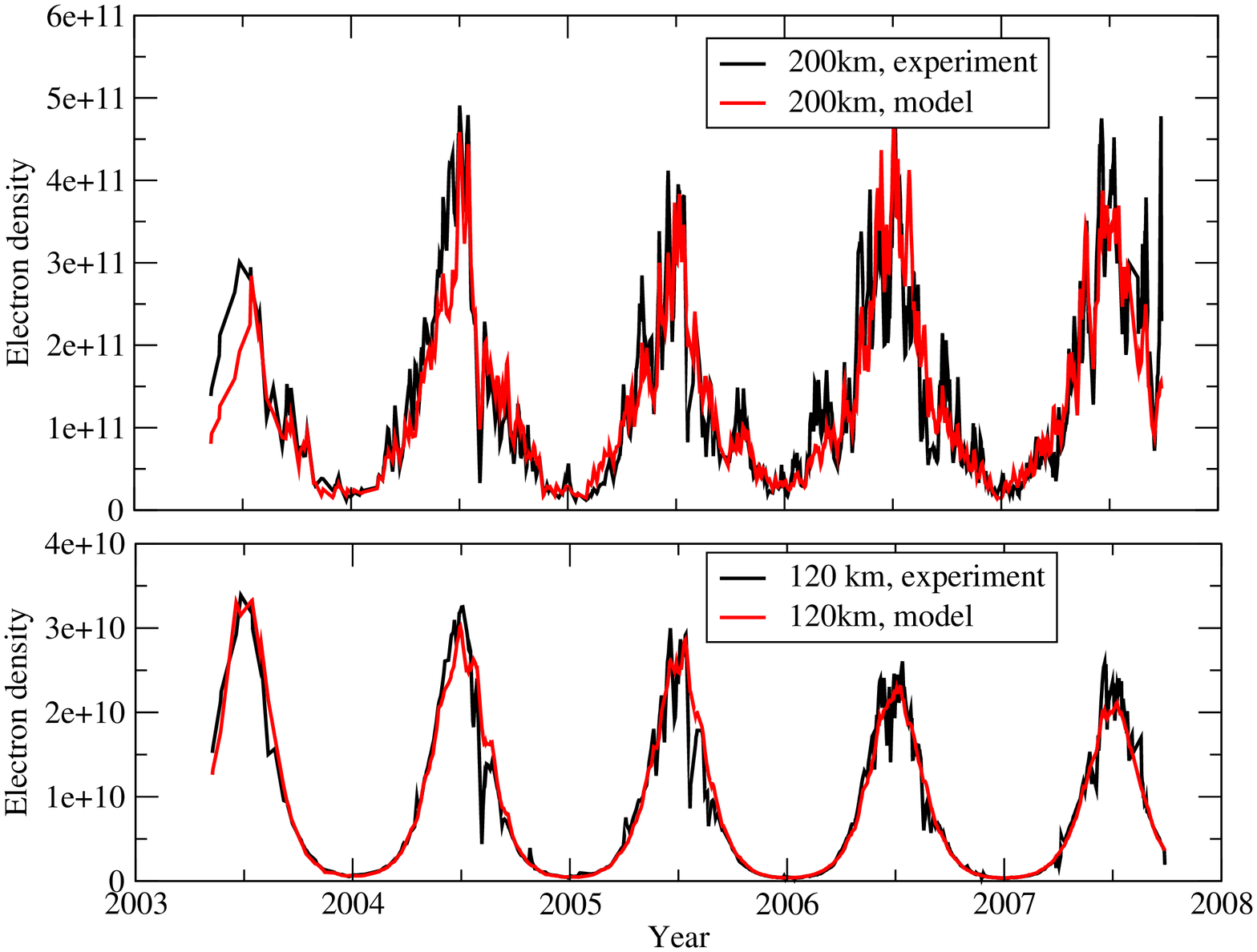}

{\small Fig 2. Comparison between the experiment and description model
for 200 km height (top) and 120km height (bottom). N=0}{\small \par}

~

At fig.3 a comparison of the experimental data with another description
model (with big number of additional periodics , $N=50$) is shown.
At the table 1 the accuracy for this model over the two different
sets - analyze and prediction ones (columns 4-5) is shown . 

\includegraphics[scale=0.5,angle=270]{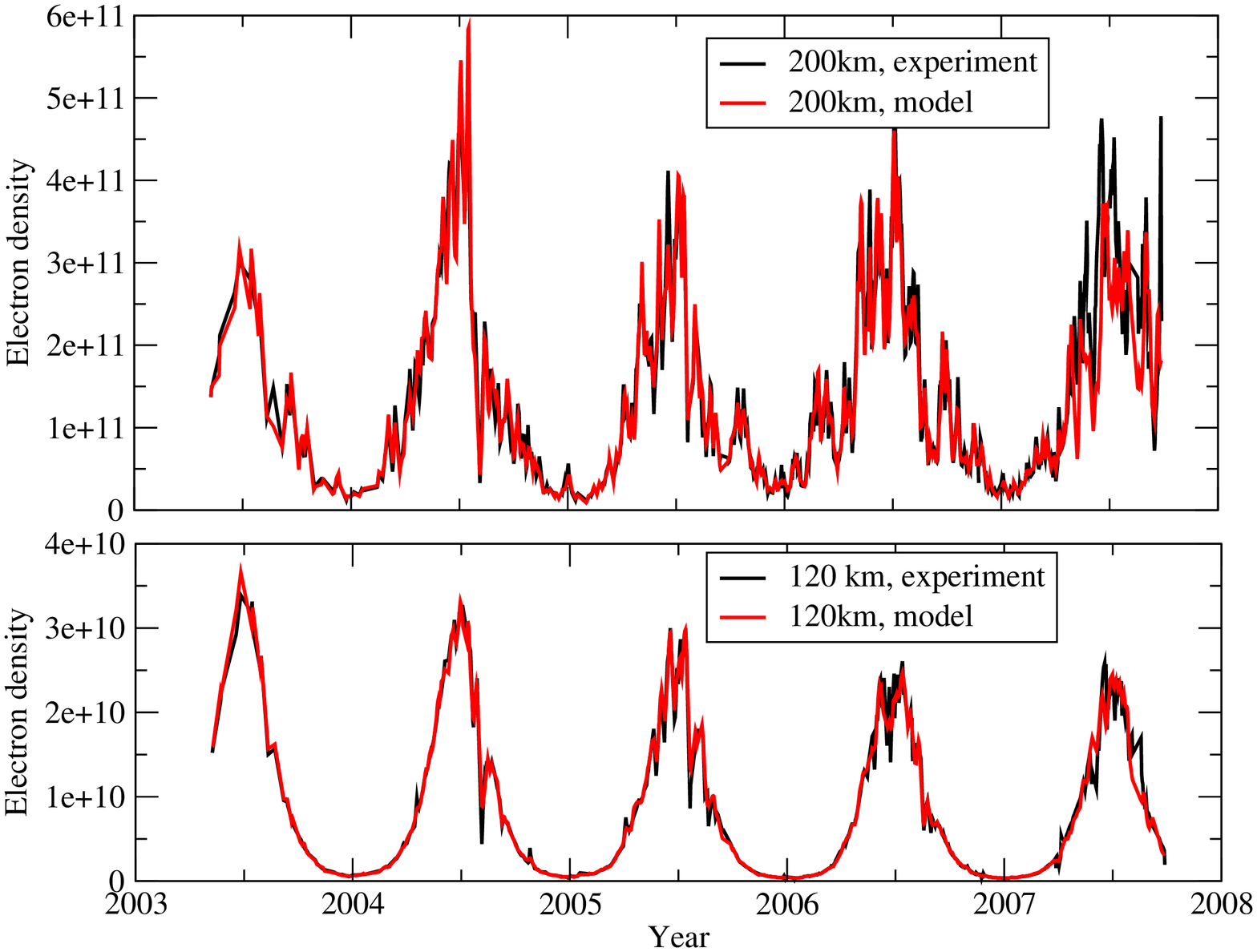}

{\small Fig 3. Comparison between the experiment and description model
for 200 km height (top) and 120km height (bottom). N=50}{\small \par}

~

At fig.4 a comparison of the experimental data with prediction model
is shown. At table 1 the accuracy for this model in two different
sets - analyze and prediction ones (columns 6-7) is shown. A number
of additional periodics is shown at column 8.

\includegraphics[scale=0.5,angle=270]{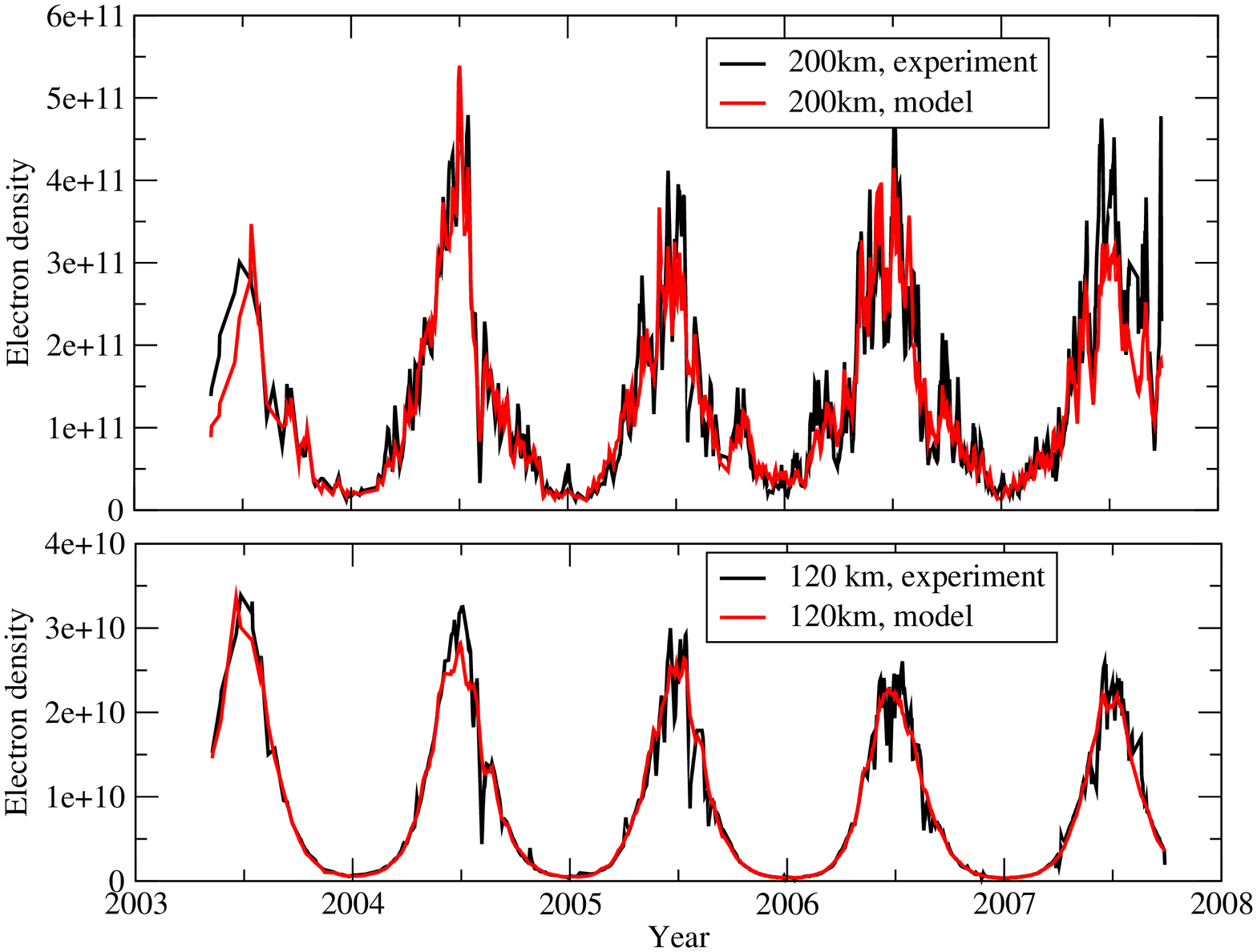}

{\small Fig 4. Comparison between the experiment and prediction model
for 200 km height (top) and 120km height (bottom).}{\small \par}

~

{\small Table 1. Model accuracy depending on model, data set and height. }{\small \par}

\begin{tabular}{|c|c|c|c|c|c|c|c|}
\hline 
Height&
analyze(0)&
preidction(0)&
analyze(50)&
prediction(50)&
analyze(opt)&
prediction(opt)&
N opt\tabularnewline
\hline
\hline 
120&
9.6\%&
11.2\%&
5.5\%&
11\%&
9.1\%&
8.8\%&
4\tabularnewline
\hline 
140&
10\%&
12.1\%&
7\%&
17\%&
10.9\%&
12.3\%&
2\tabularnewline
\hline 
160&
14.4\%&
9.3\%&
7.6\%&
19\%&
12.1\%&
12.3\%&
4\tabularnewline
\hline 
180&
19.8\%&
10.8\%&
8.5\%&
24\%&
15.8\%&
15.8\%&
5\tabularnewline
\hline 
200&
25.4\%&
20\%&
10.7\%&
32.5\%&
22.8\%&
23.2\%&
3\tabularnewline
\hline 
220&
24\%&
14.8\%&
10.2\%&
26.6\%&
19.2\%&
21.3\%&
6\tabularnewline
\hline 
240&
22.1\%&
14.5\%&
9.2\%&
23.8\%&
17\%&
19.4\%&
7\tabularnewline
\hline
\end{tabular}

{\small Column 1. - Height, km; Column 2. - descirption model, error
over analyze set(2003-2006 years),N=0; Column 3. - descirption model,
error over prediction set(2007 year),N=0; Column 4. - descirption
model, error over analyze set(2003-2006 years),N=50; Column 5. - descirption
model, error over prediction set(2007 year),N=50; Column 6. - prediction
model, error over analyze set(2003-2006 years),N from stop condition;
Column 7. - prediction model, error over prediction set(2007 year),N
from stop condition; Column 8. Number of additional periodics for
prediction model.}{\small \par}

~

From the table 1 one can see that description model provides a better
accuracy over analyze set than prediction model over analyze set (columns
4,6). 

From the data also one can see that accuracy of description model
is increased with increasing number of parameters (columns 2,4). But,
over prediction set of the data (not included into analysis) the prediction
model could make better accuracy than description model with big number
of parameters (columns 5,7). So when making a model for good description
of the ionospheric electron density variations we should take an exact
set of parameters. Both small number of parameters and big number
of parameters could increase the error when using it over prediction
set of the data, i.e. when using the model for prediction purposes.

From the analysis one could also see that with increasing of the height
the number of additional periodical variations that should be taken
into account (column 8) increases.

\section{Conclusion}

In the paper the step-by-step principles for making local model of
electron density (\ref{eq:19}) are described. They are based on modulation
principle - electron density dependence on time is a product of temporal
variations caused by different sources: solar radiation, magnetic
activity, Earth orientation and unknown additional periodical processes
(but not a sum, as sometimes suppose when making local models). A
multiranges modulation principle is also described, that allows automatically
extend set of parameters by using new indexes obtained by filtration
(or averaging) basic set of parameters over the time. In the paper
we have described two approaches to the model creation - descriptional
and predictional ones.

To test the approach a three different models were created for daily
electron density logarithm using the stated principles. We have used
the data of Irkutsk digisonde over the period 2003-2007 years for
testing. It becomes clear that non-optimal choice of the number of
model parameters could increase prediction error, inspite the error
over the set, used for analysis, will decrease. It is shown that one
year prediction has accuracy about 9-23\% depending on the height,
and the highest error corresponds to the height about 200km. From
the modelling we can also see that with increasing of the height the
number of parameters increases, and this could be caused by inaccuracy
of the model or by not including an additional physical mechanisms
into consideration.

Author thanks to L.A.Schepkin, N.V.Ilyin and K.G.Ratovsky for useful
discussions, and to A.A.Berngardt for help in preparing this publication.

The work was done under support of RFBR grant \#05-07-90212-v


\begin{thebibliography}{Gonorovskij,1986}
\bibitem[Reinisch,1997]{Reinisch_1997}Reinisch, B.W., Haines, D.M.,
Bibl, K., Galkin, I., Huang, X., Kitrosser,D.F., Sales, G.S., and
Scali, J.L. Ionospheric sounding support of OTH radar //Radio Sci.
32 (4), 1681-1694, 1997.

\bibitem[Holt et al,2002]{Holt_et_al_2002}Holt J.M., Shun-Rong Zhang,
Buonsanto M.J., Regional and local ionospheric models based on Millstone
Hill incoherent scatter radar data //Geoph.Res.Letters, 29, 8, 1207,
doi:10.1029/2002GL014678, 2002

\bibitem[Hargreaves,1992]{Hargreaves_1992}Hargreaves, J. K., The
Upper Atmosphere and Solar-Terrestrial Relations. Cambridge University
Press, 1992

\bibitem[Chapmen,1931]{Chapmen_1931}Chapman S. The absorbpion and
dissociative of ionizing effect of monochromatic radiation in an atmosphere
on a rotation Earth //Proc.Phys.Soc., 1931, 43, 26, p.483. 

\bibitem[Shimazaki,1959]{Shimazaki_1959}Shimazaki T. Theoretical
study of the dynamical structure of the ionosphere //Journ.Radio Res.
Labs., 1959, 6, 24, pp.109-242.

\bibitem[Kuklin et al,2000]{Kuklin_et_al_2000}Kuklin G.V., Orlov
I.I., Berngardt O.I., On technique for determining periodics based
on Wolf number analysis //Issledovania po geomagnitizmu,aeronomii
i fizike solnca, Izdatelstvo SB RAS, 2000, 110, pp.7-12 (in russian). 

\bibitem[Gonorovskij,1986]{Gonorovskij_1986}Gonorovskij I.S. Radiotechnical
circuites and signals. Radio i Svjaz, 1986, 4-th edition. (in russian)
\end{thebibliography}
\end{document}